\documentclass{article}[10pt]
\usepackage{epsfig}

\newcommand{\lsim}{\lower .5ex\hbox{$\buildrel < \over {\sim}$}}
\newcommand{\gsim}{\lower .5ex\hbox{$\buildrel > \over {\sim}$}}

\newcommand{\ndt}{\noindent}

\newcommand{\ov}{\overline}

\begin{document}
% \maketitle

\begin{flushright}
	DFUB 00-20
\end{flushright}
\vskip 2. cm
\begin{center}
{\bf\Large Hadron-hadron and hadron-nuclei
collisions at high energies}
\vskip 0.4 cm
{\bf \large Invited paper at Chacaltaya Meeting on Cosmic Ray Physics, \\
Lapaz,
23-27 July, 2000\\}
\vskip 0.4 cm
{\large\noindent G. Giacomelli and R. Giacomelli\\
\vskip 0.2 cm
Dipartimento di Fisica dell'Universit\`a di Bologna 
and INFN, I-40127 Bologna, Italy}
\end{center}

\begin{abstract}
A brief review is made of the present situation of hadron-hadron and
hadron-nuclei
total, elastic and inelastic cross sections at high energies.
\end{abstract}
\section {Introduction}

 In the 1960's-1980's high quality secondary beams became available at 
proton synchrotrons of increasing energies (PS, AGS, IHEP, FNAL, SPS). The 
secondary charged hadron beams contained the six stable or quasi stable
charged particles $\pi^{\pm}$, K$^{\pm}$, $p$, $\ov p$. 
The main experimental lines of research were: i) total hadron--hadron
 ($hh$)
 cross section measurements at high energies ($p_{lab} > 20$  GeV/c);
 ii) elastic $hh$ scattering measurements; iii) measurements of absorption 
cross sections of hadrons in nuclei (as byproducts of $hh$ total cross section
measurements);
iv) hadron production at forward angles in $p-nuclei$ collisions, and 
v) inelastic $hh$ collisions [1-5]. In some high intensity beams many
 particle searches
were made, which also lead to the study of $\ov d,~\ov t$ and $\ov {He}_3$
production [6]. \par

In order to reach higher energies it became necessary to build hadron
colliders. The first was the ISR $pp$ and ${p\ov p}$ collider at CERN, which
allowed to reach c.m. energies between 22 and 63 GeV; then the $Sp\ov pS$
collider at CERN
allowed ${p\ov p}$ collisions from 600 to 900 GeV; finally the Fermilab
Tevatron collider allowed $p\ov p$ collisions up to 1.8 TeV.

 The
first experiments performed at the ISR were relatively simple
dedicated experiments,
like those for total cross section and elastic scattering measurements, and
single arm
spectrometers for the study of inelastic collisions [7-8]. Then followed
general purpose
detectors: the SFM  at the ISR [9], the general
 purpose detectors  UA1 and UA2 and experiment
 UA4 at the $Sp\ov pS$
collider, and finally the
general purpose detectors CDF, D0 and  the specialized experiments E710, F8 at
 Fermilab
[10].\par 

The highest energies were and are still obtainable only with Cosmic Rays.
We shall discuss total cross sections at high energies, the high 
energy low$-p_{t}$ parameters and some features of  inelastic collisions
[11, 12].\par

\section {Hadron-hadron total cross-sections}

At fixed target accelerators (AGS, IHEP, FNAL) the total cross sections
were measured with the transmission method, with relative precisions smaller
 than 
$1\%$ and systematic scale errors of about $2\%$.
The measurements of the total cross sections at the $pp$ and $\ov pp$
colliders
required the development of new experimental techniques: the scattering
of particles was measured at very small angles, with detectors positioned in
reentrant containers (\lq\lq Roman pots") located very close to the
 circulating beams. The combination of statistical and systematic 
uncertainties are  $\geq 10 \%$. Fig 1 summarizes the present status
of high energy total cross section measurements: for
$E_{cm} = \sqrt{s} > 3.4$ GeV all
 total $hh$
cross sections decrease, reach a minimum and then increase with increasing 
energy (the $K^{+}p$ total cross section was already 
increasing at Serpukhov energies [1]). Moreover the differences between the cross
 sections $\ov x p$ and $x p$ decrease with increasing energy [5]. 

There is not a unique interpretation for this
 rise, though in many QCD inspired models it seems to be connected with
 the increase of the number of minijets and thus to semi-hard gluon  
interactions.

Most of the high energy elastic and total cross section data have been usually
interpreted in terms of Regge Poles, and thus in terms of Pomeron exchange.
 Even if the Pomeron was introduced  long time ago we do not have a
 consensus on its exact definition and on its detailed substructure. Some 
authors view it as a "gluon ladder".

Future experiments on hadron-hadron total cross sections remain 
 centered at the Fermilab Collider (for $p\ov p$). The near
future will rely on the BNL-RICH Collider, and later, on the LHC proton-proton
and heavy ion
Collider at CERN. Large area cosmic ray experiments may be able to improve
the data in the ultra high energy region [12].

\begin{figure}[htb]
%\vspace{8cm}
\begin{center}
\mbox{
\epsfig{figure=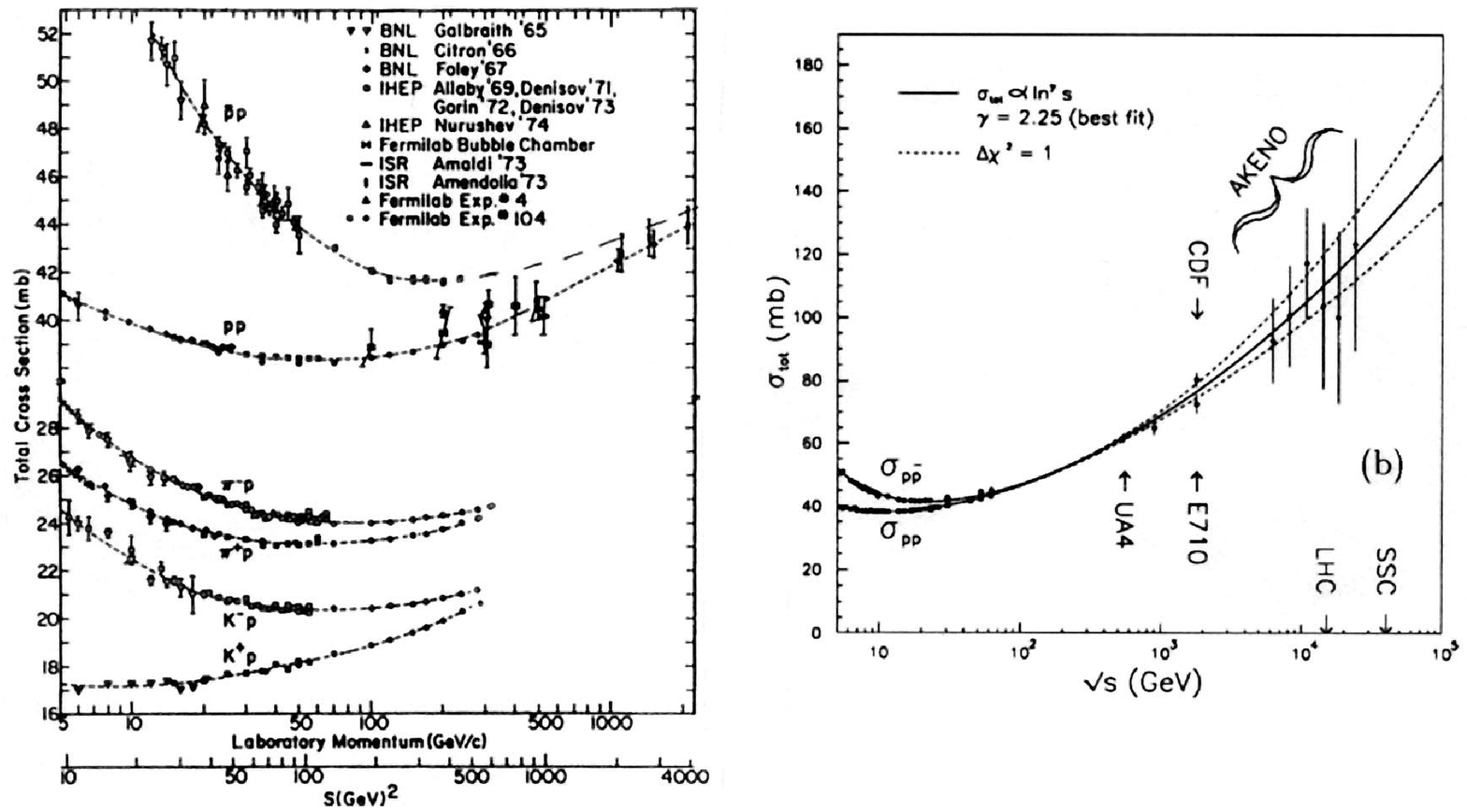,width=14cm,height=8cm}}
\end{center}
\vspace{0.3cm}
\caption{\small Compilation of total cross-sections (a) for high-energy
$hh$ scattering and
(b) for higher energy $\ov pp$ and $pp$, 
 including cosmic ray measurements; the solid line is a fit;
the uncertainty region  is delimited by dashed lines.}
\label{1}
\end{figure}

\section{ Hadron-hadron elastic scattering}

The differential cross section for the elastic
scattering of unpolarized particles from unpolarized targets has a simple
structure 
in the  high energy region.  It depends on two variables: the energy
and an angular variable,
which is usually chosen to be the square of the four momentum transfer $t$.
The energy dependence is of the $ln~s$ type.
The angular distribution may be divided in four regions:
i) The Coulomb region for $|t|<0.001$ (GeV/c)$^2$;
ii) The Coulomb-nuclear interference region for
$0.001<|t|<0.01$ (GeV/c)$^2$; measurements in this region yield information
on
the ratio $\rho$ of the real to the imaginary part of the forward scattering
amplitude.
iii) The nuclear diffraction region proper for $0.01<|t|<0.5$ (GeV/c)$^2$;
here the most important parameter is the slope B, or the slopes $b_i$, of
the
diffraction pattern.
iv) The large angle region for $|t|>0.5$ (GeV/c)$^2$:
for $E_{cm}\leq 100$ GeV
it is characterized by a dip-bump structure which
resembles that from diffraction from an opaque disc.
Most of these features are observed in the $pp$ and $\ov pp$ elastic
scattering
angular distributions at $E_{cm}= 53$, 546 and 1400 GeV, see
Fig. 2.\par

\begin{figure}[htb]
%\vspace{5cm}
\begin{center}
\mbox{\epsfig{figure=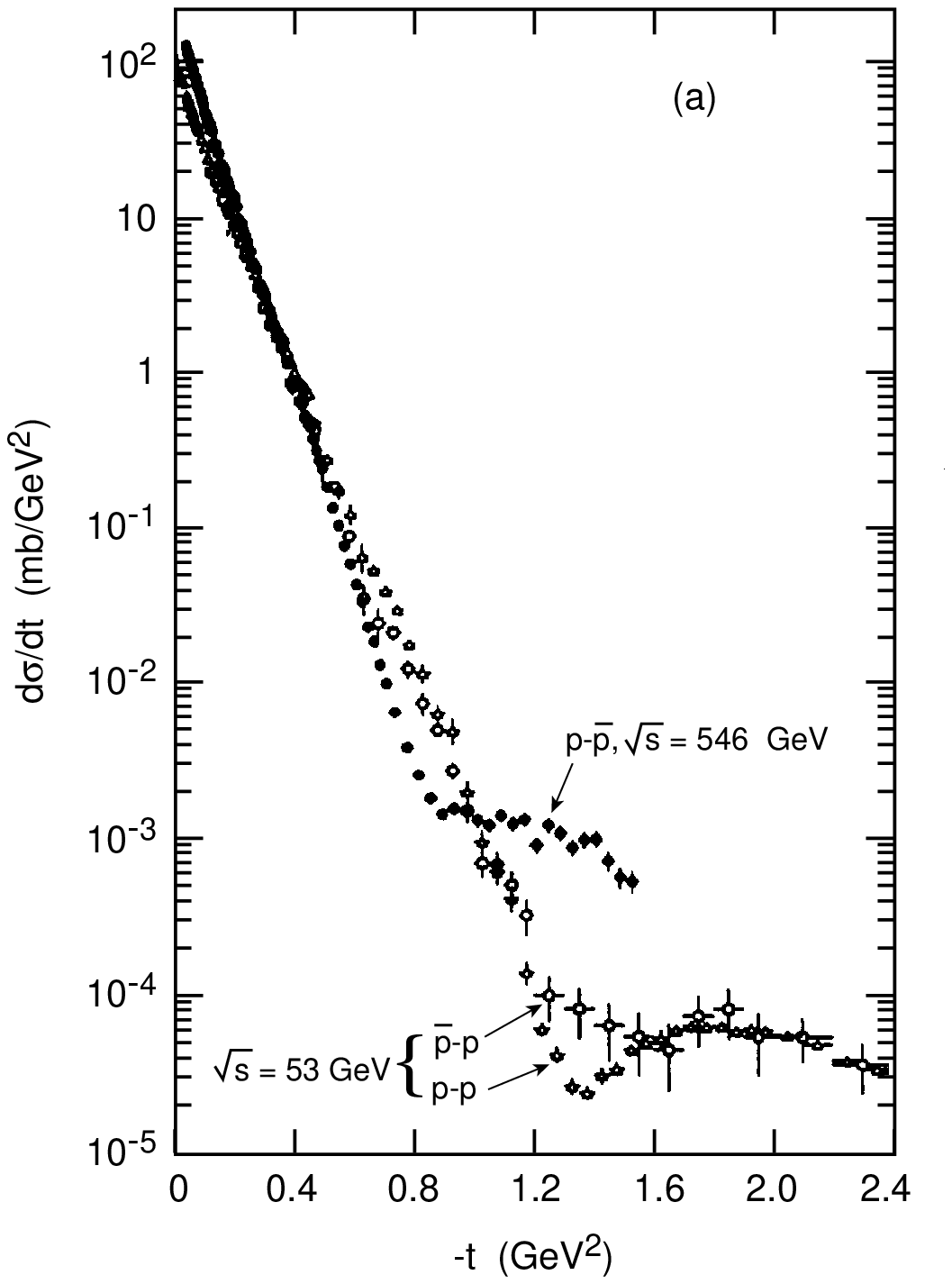,width=6.3cm,height=6.3cm}}
%\hskip 0.5 truecm
\mbox{
\epsfig{figure=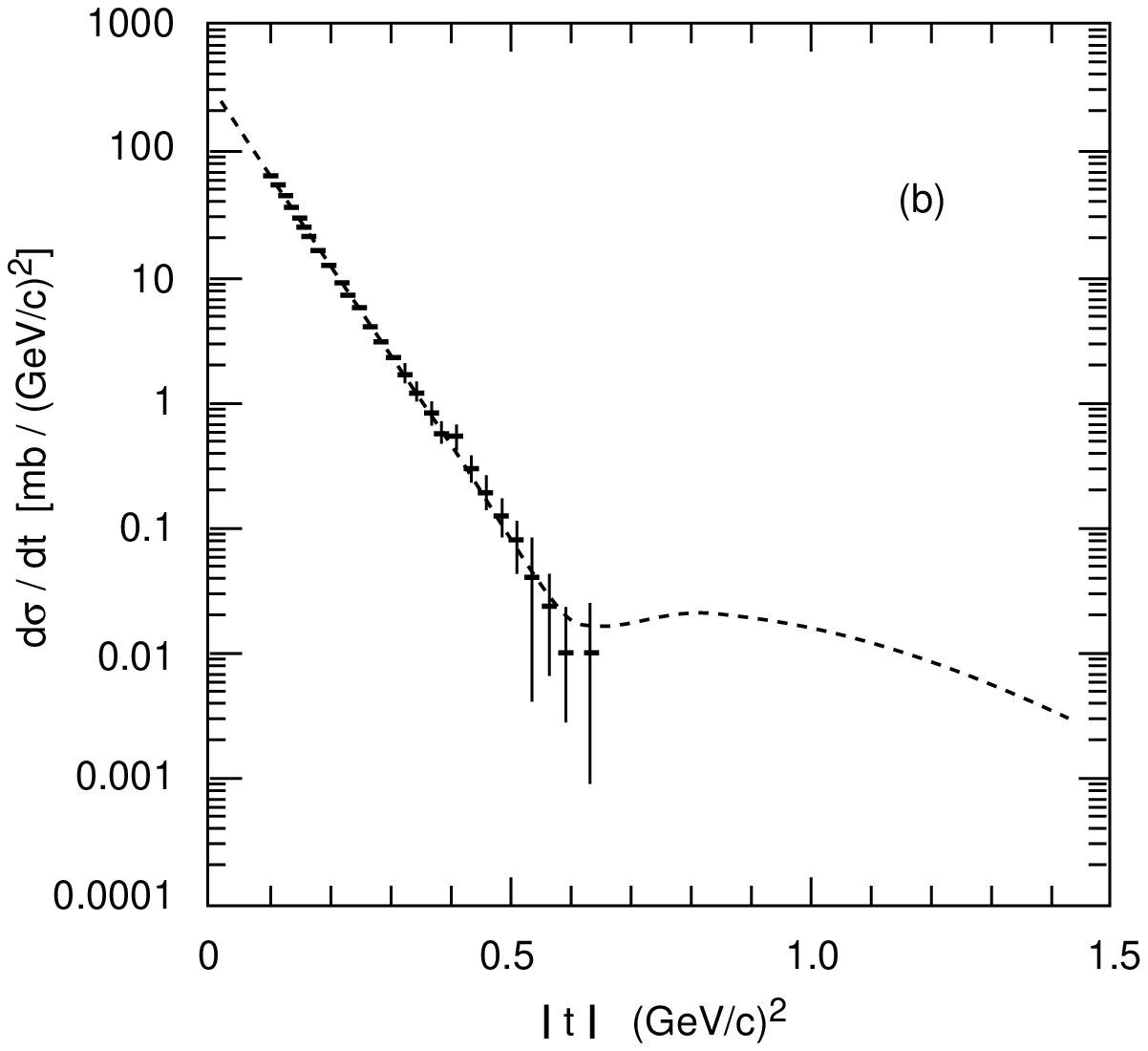,width=6.3cm,height=6.3cm}}
\end{center}
\caption{\small Differential elastic  cross sections for $pp$ and
$\ov pp$ (a) at 53 and 546 GeV, (b) at 1.8 TeV.}
\label{fi:2}
\end{figure}

One usually defines the following high energy scattering parameters: i) the
total cross section, $\sigma_{t}$; ii) the slope $B$ of the differential elastic
nuclear cross section 
$d \sigma/dt = Ae^{Bt}$, at $|t|\simeq 0.1$ (GeV/c)$^2$; iii) the ratio
$\rho = (ReA/ImA)_{t=0}$;
iv) the opacity  $O=2\sigma_{el}/\sigma_{t}$.
At very high energies $\sigma_t$,  $B$ and $O$
rise with energy, while $\rho$ is slightly positive. From the behaviour of
these
parameters emerges the picture of a proton which becomes bigger (larger $B$)
and more opaque (larger $O$) as the energy increases.
At the highest energy $O\simeq 0.48$, which is still smaller than
the black disk value of 1.\\

\section{Hadron-hadron inelastic processes}

Fig. 3 shows pictorially various inelastic processes at low $p_{t}$: single
and double diffraction dissociation and inelastic processes; these are the 
dominant processes with the largest cross sections, concentrated in particular
in the very forward angular region:
this is the region which is most important for cosmic ray experiments.

\begin{figure}[htb]
%\vspace{5cm}
\begin{center}
\mbox{\epsfig{figure=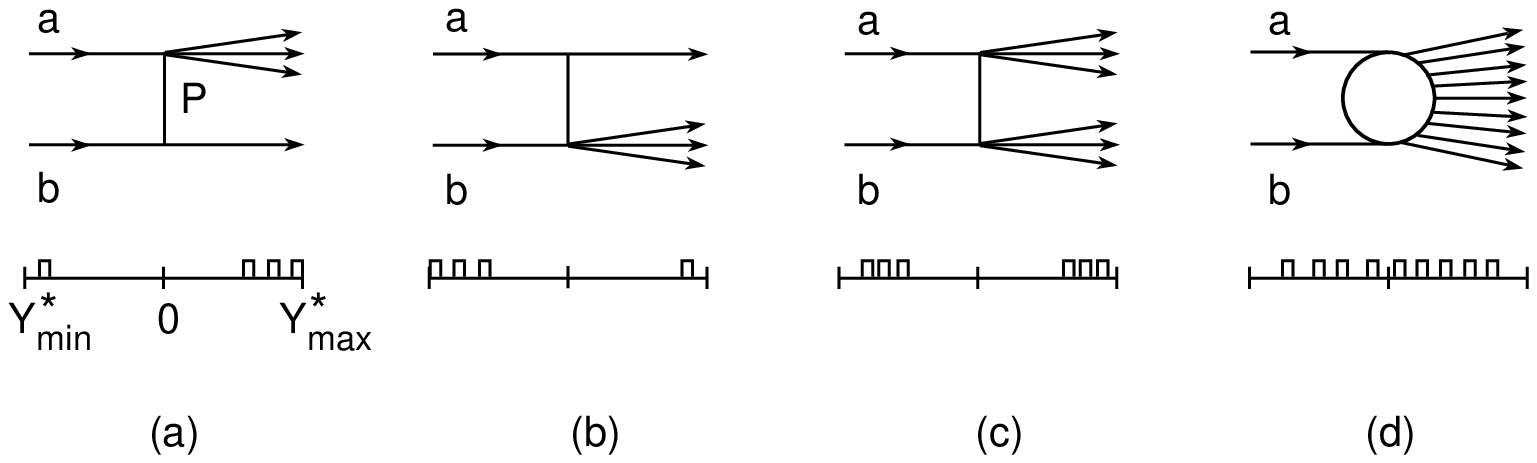,width=13cm,height=3cm}}
\end{center}
\caption{\small
Pictorial description of inelastic, small $p_{t}$  processes with
characteristic rapidity distributions. P indicates Pomeron exchange.
(a) Fragmentation of the beam particle $a$; (b) fragmentation of the target
$b$;
(c) double fragmentation of a and b; (d) central collision.}
\label{3}
\end{figure}

At high energies the total $\ov pp$ cross section may be written
as
$$
\sigma_{t}=\sigma_{el}+\sigma_{inel}=\sigma_{el}+\sigma_{sd}+\ov\sigma_{sd
}
+\sigma_{dd}+\sigma_{nd}\eqno (1)$$
\ndt where $\sigma_{el}$ is the elastic cross section, $\sigma_{sd}$ is the
single
 diffractive cross section when the incoming proton fragments into a number
of
particles, $\ov\sigma_{sd}$ is the single diffractive cross section for the
fragmentation of the antiproton (at high energies
$\sigma_{sd}=\ov\sigma_{sd}$). $\sigma_{dd}$ is the double diffractive cross
section, $\sigma_{nd}$ is the non-diffractive part of the inelastic cross
section (Fig. 3).
The elastic, single diffractive and double diffractive processes give rise
to
low multiplicity events with particles emitted in the very forward
 region in the c.m. system. The non-diffractive cross section is
the
main part of the inelastic cross section; it gives
rise to high multipliciy events and to particles emitted at all angles. Most
of the non-diffractive cross section concerns particles emitted with low
transverse momentum ({\it low $p_t$ physics}) with properties which change
slowly
with c.m. energy ({\it $ln~ s$  physics)}. A  small part of the
non-diffractive cross section is due to central collisions among the
colliding
particles and gives rise to high $p_t$ jets of particles emitted at
relatively
large angles ({\it large $p_t$ physics)}. The contribution of jet physics
increases with c.m. energy.\par

In Fig. 4a are shown, vs $\sqrt{s}$, the  average charged multiplicities in
$pp$ and $\ov pp$ collisions. The data may be fitted to a power law of
$ln~s$:
$$ \ov n=A+B~ln~s+C~ln^2~s\simeq 3.6-0.45~ln~s+0.20~ln^2~s \eqno   (2)$$
\ndt At $\sqrt{s}=1.8$ TeV are produced on average 40 charged
and 20 neutral particles.
Fig. 4b shows the average number of $\pi^{\pm}$, $K^{\pm}$, $p$ and $\ov p$
produced in $pp$ collisions up to $\sqrt{s}$ $\simeq 100$ GeV. Pions
are the dominantly produced hadrons.
\begin{figure}[htb]
%\vspace{7cm}
\begin{center}
\mbox{
\epsfig{file=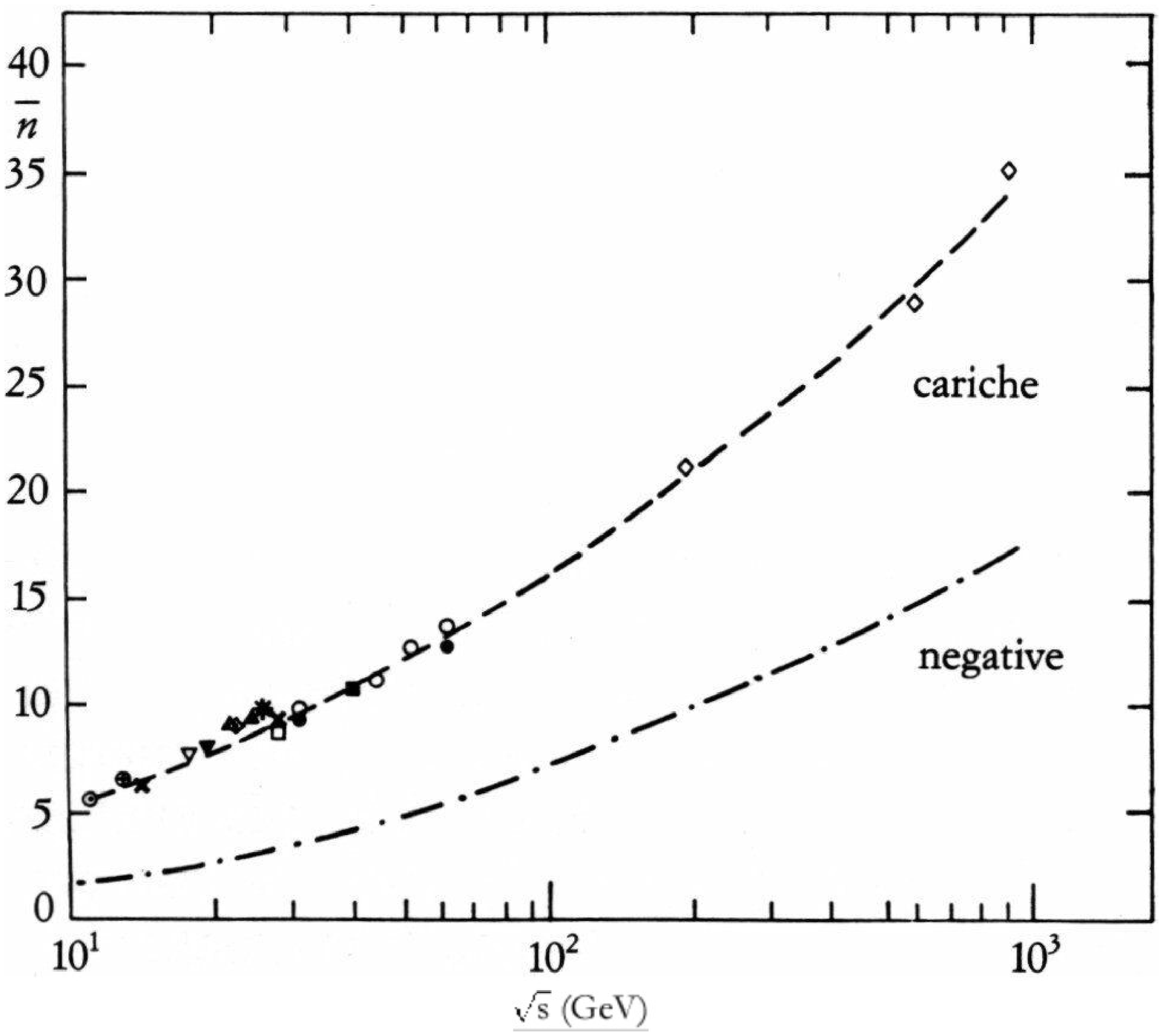,height=6cm}
\epsfig{file=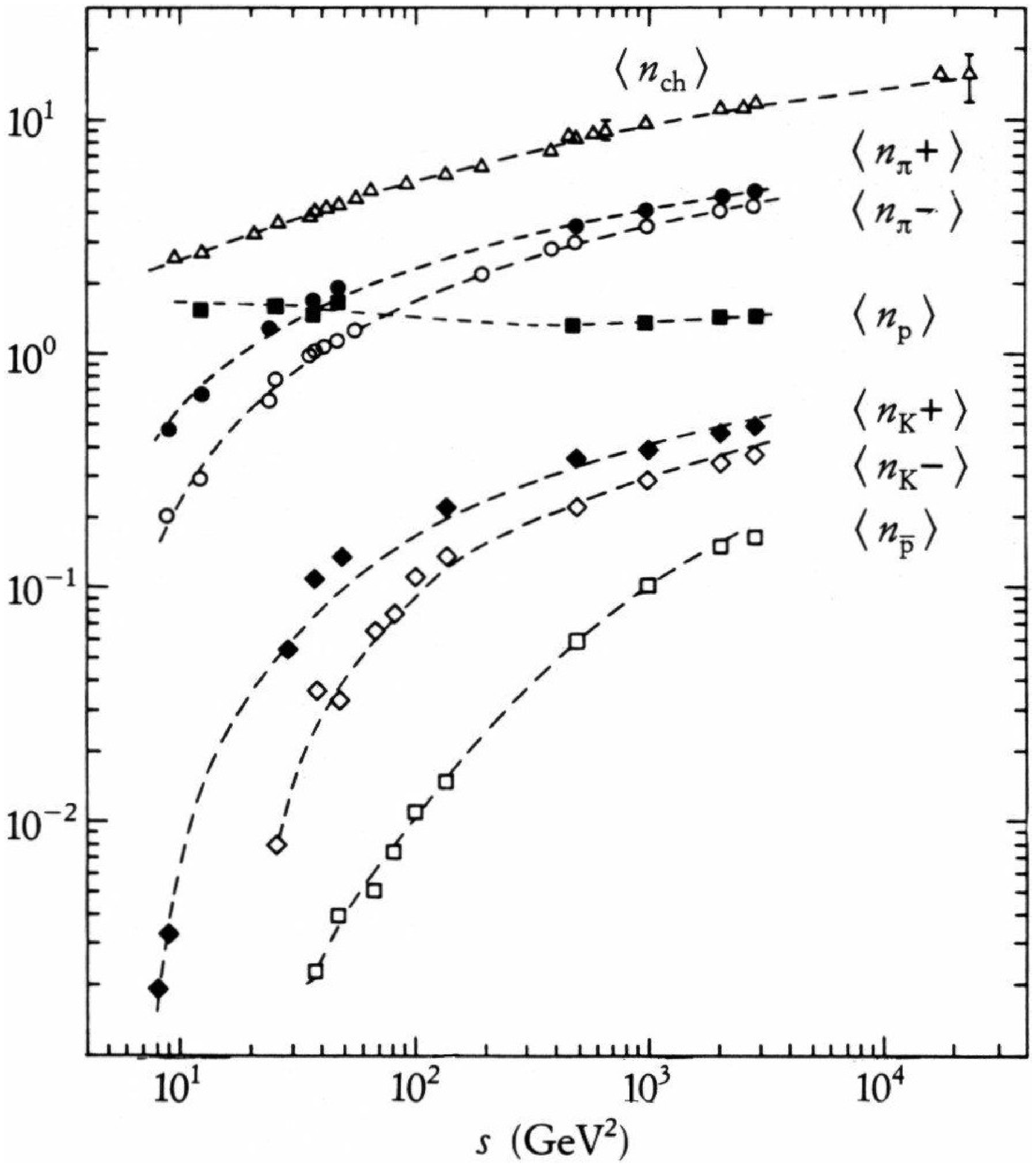,height=6.5cm}
}
\end{center}
\caption{\small (a) Average charged  multiplicities $\ov n$ for $pp$ and
$\ov pp$ collisions vs $E_{cm} = \sqrt{s}$. (b) Average number of
$\pi^{\pm}$ $K^{\pm}$, $p$ and $\ov p$ produced in $pp$ collisions up to
$E_{cm}= 63$ GeV.}
 \label{4}
\end{figure}
Hadrons are mainly produced at low transverse momenta. The average $p_t$
increases slowly with $\sqrt{s}$: it is $\langle p_{t}\rangle \simeq 0.36$ GeV/c
in  $20<\sqrt{s} <100$ GeV; then it increases slowly up to $\simeq 0.46$
GeV/c at $\sqrt{s}=1.8$ TeV. The simplest interpretation of these
features is in terms of thermodynamic models:  hadrons are emitted from a
region with a temperature  $T\simeq 130-200$ MeV.
QCD, at present cannot be used to calculate
the data at low $p_t$, because  the strong coupling constant is
large and  perturbative methods cannot be used. Therefore one has to use
models. Hard collisions can instead be  calculated by perturbative QCD.

\section{Hadron-nucleus and nucleus-nucleus collisions}

Usually, as byproducts of  $hh$ total cross section measurements,
the absorption cross sections
of $\pi^{\pm}$, $K^{\pm}$, $p$ and $\ov p$ on various nuclei (Li, C, Al, Cu,
Sn, and Pb) were
measured at incident lab momenta up to 280 GeV/c [2, 4, 11]. Most
absorption cross
sections decrease slowly for p$_{lab}$ up to $\simeq$  50 GeV/c; for
higher p$_{lab}$ they are
almost constant. 

The data at each energy were fitted to the simple  expression
$$ \sigma_a (A) = \sigma_0 A^{\alpha} \eqno (3) $$
\ndt where $A$ is the atomic weight of the target nucleus. Examples
are shown in Fig. 5a at three lab momenta for three different incoming
hadrons.
For all incident particles except antiprotons, the value of
$\sigma_0$  increases by up to 10\% as the incident p$_{lab}$ increases from
60 to
280 GeV/c, with the largest increase for $K^+$. For $\ov p$, $\sigma_0$ 
decreases with increasing p$_{lab}$.
 Fig. 5b shows the parameters $\sigma_0$ and  $\alpha$
 vs the
corresponding $hp$ total cross section:
$\sigma_0$  rises monotonically
with $\sigma_{hp}$; the values of $\alpha$ are consistent with $\alpha$
approaching  0.67 for large values
of $\sigma_{hp}$ as would be expected for an opaque nucleus.\par

\begin{figure}[htb]
%\vspace{8cm}
\begin{center}
\mbox{
\epsfig{figure=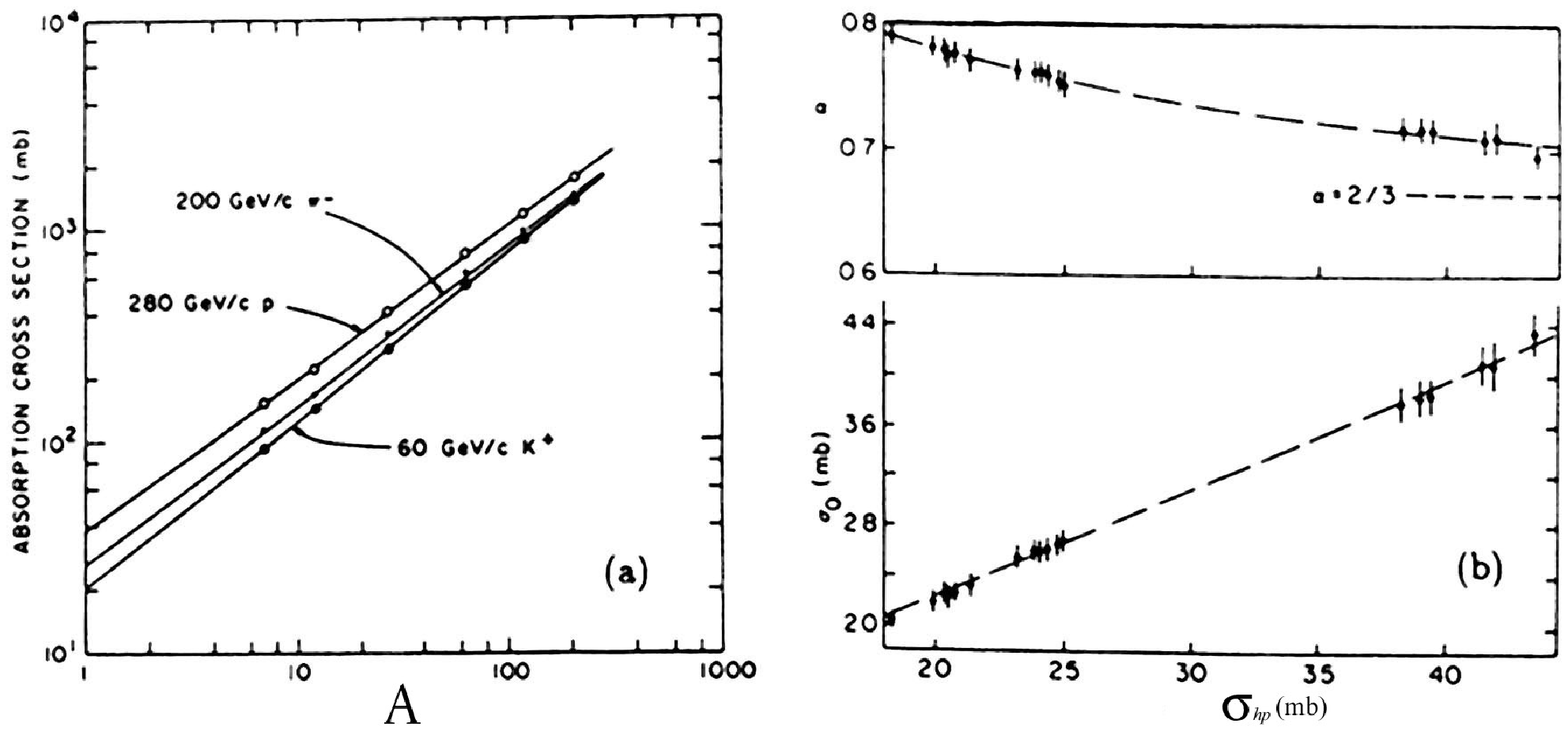,width=14cm}
}
\end{center}
\caption{\small (a) Absorption cross sections vs atomic weight; the
solid lines are fits to Eq. 3.
(b) The parameters $\sigma_0$ and $\alpha$ vs
the corresponding $hp$ total cross sections, $\sigma_{hp}$.}
\label{5}
\end{figure}

The fragmentation cross sections of various nuclei were measured at
many energies using  nuclear track detectors, which 
yield high resolution measurements of the restricted energy loss.
This arises from the relatively low energy required to break
the polymeric bonds of the detectors and because
fluctuations due to  energetic $\delta$-electrons do not
contribute to the latent track formation. The radiation damage along the
path of
an incoming nucleus may be developed to microscope-visible cones by
chemical etching. Fig. 6a shows the charge distribution
obtained with 200 GeV/nucleon S$^{16+}$
ions and their fragments produced in a Cu target [13]: notice the
good charge resolution and the absence of nuclei with fractional charges.
The cross sections in
Fig. 6b  are relative to fragments of 158 GeV/nucleon Pb$^{82+}$ ions in
a target with $\ov A = 11.5$
with a variation of atomic number  $\Delta Z= 1-7$ with respect to the
$Z=82$
of the incoming ions.

\begin{figure}[htb]
%\vspace{5.cm}
\begin{center}
\mbox{
\epsfig{figure=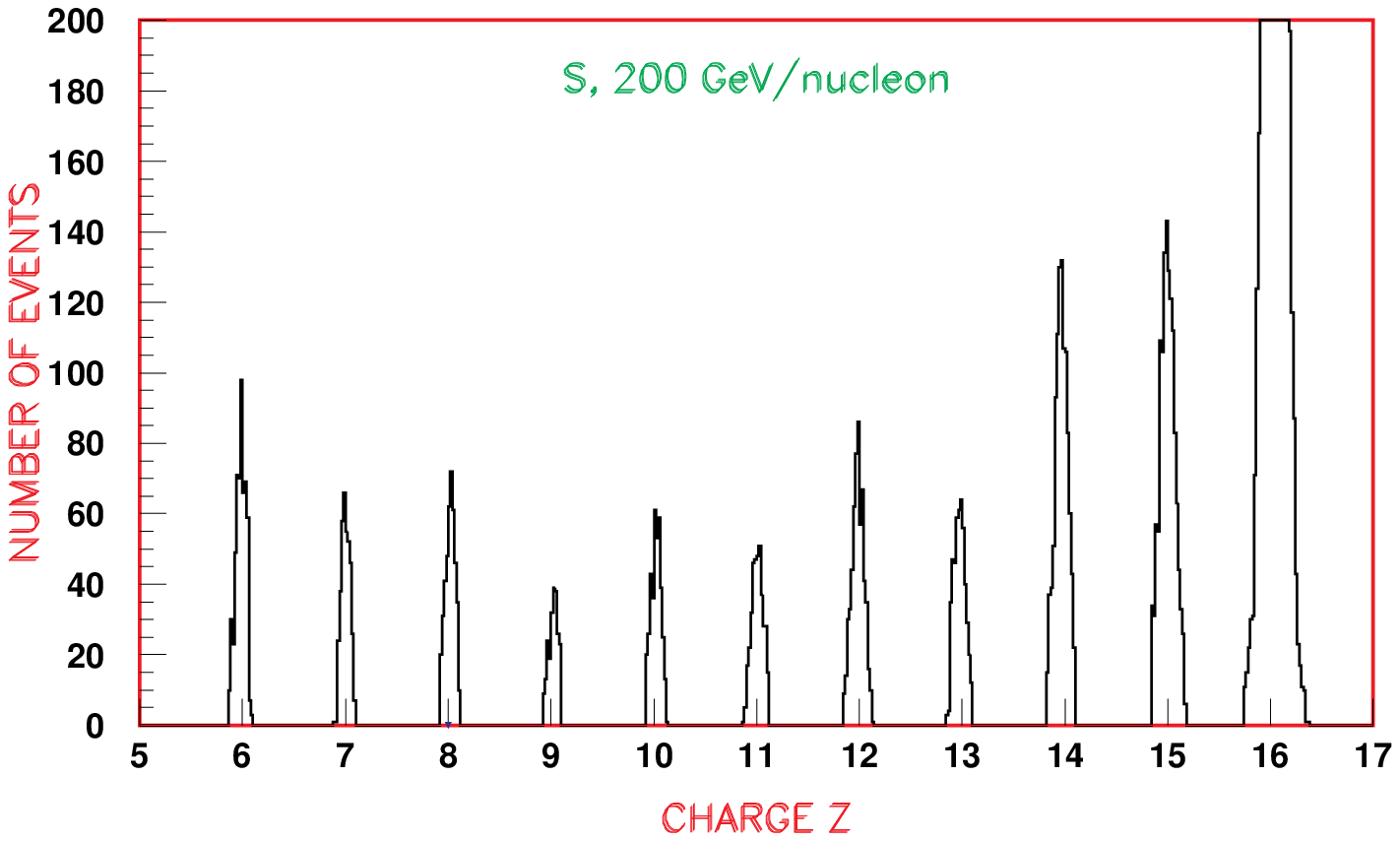,width=7.5cm}
\epsfig{figure=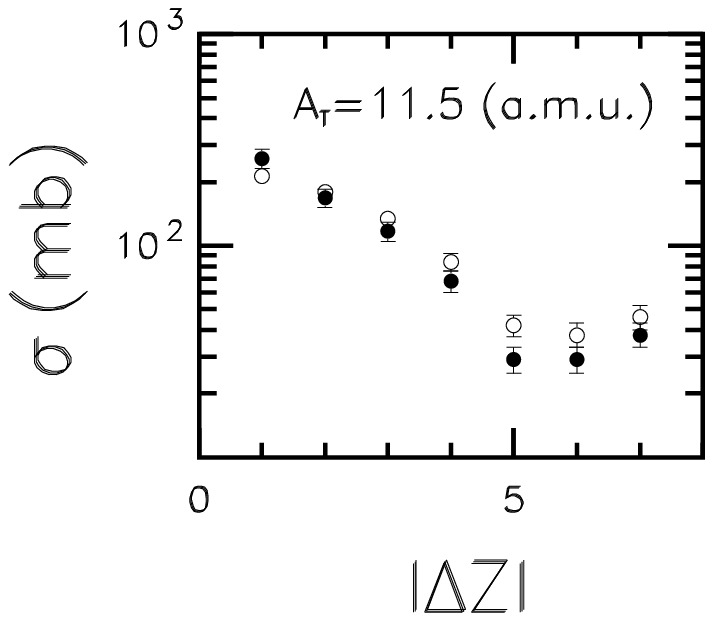,width=5.5cm}
}
\end{center}
\caption{\small (a) Charge distribution of 200 GeV/nucleon S$^{16+}$
ions
and their fragments [13].
(b) Cross sections in a target $\ov A = 11.5$ of 168 GeV/nucleon
Pb$^{82+}$ ions
into nuclear fragments for $\Delta Z$ ranging from 1 to 7;
the open and black points refer to two different computing methods.
}
\label{6}
\end{figure}

\section{Conclusions. Perspectives}
A wealth of experimental information has been obtained
since the early 1960's on high-energy
hadron-hadron and hadron-nucleus cross sections, starting with simple
equipments
and proceeding to more complex apparata and better beams. 

The higher energy data coming from collider experiments have large systematic
uncertainties, mainly connected to the poor knowledge of the collider luminosity,
the difficulty of measuring with some precision the single and double
diffraction cross sections, etc.

Most of the
information on hadron-hadron collisions 
was interpreted in terms of phenomenological models.
Much more work
is needed to interpret and systemize the available data, since only the
 small
part of the cross section corresponding to large transverse
momenta may be interpreted in terms of QCD.

Few and not too precise data are available at very low $p_{t}$
and very small angles, which is the 
region where more interest lies for cosmic rays.
The information is here codified in Monte Carlos, which
are becoming progressively more complicated. The Monte Carlos include 
hadron-nucleus collisions [for which the data are much less abundant than for
 $hh$ collisions]; hadron-nucleus effects are mainly interpreted in the context
of the Gluber model [14].

Nucleus-nucleus data are even less abundant, and the Monte Carlos have extra 
complications coming from the used models [14].

In the near future the RICH collider at BNL should provide data
on nucleus-nucleus collisions. In the future the different experiments
at the LHC collider should provide data on $hh$, hadron-nuclei and nuclei-nuclei
collisions at much higher energies.

We thank many colleagues for their collaboration, for many discussions and for 
explanations of Cosmic Ray Monte Carlos.
We thank Ms L. De Angelis for typing the manuscript.

\section{References}

\begin{enumerate}

\item{J. V. Allaby et al.,
% (Total cross sections of $\pi^{-}, K^{-}$ and $\ov p$
%on $p$ and d in the range 20-65GeV/c)
 Phys. Lett. 30B (1969) 500; S. P. Denisov et al.
% (Total cross sections of $\pi^{+}, K^{+}$ and p on p and d in the 
%momentum range 15-60 GeV/c)
 Phys. Lett. 36B (1971) 415; Yad Fis 14 (1971) 998.} \vspace{-0.3cm}
\item{F. Binon et al.,
% (Absorption cross sections of 25 GeV/c d in Li, C, Al, Cu,Pb)
 Phys. Lett. 31B (1970) 230.} \vspace{-0.3cm}
\item{W. F. Baker et al.,
% (Measurement of $\pi^{+}, \pi^{-}$,
% $K^{+}, K^{-}, p$ and $\ov p$ production by 200 and 300 GeV/c protons) 
Nucl. Phys. B51 (1974) 303;
A. S. Carroll et al.,
% (Total cross sections of $p and \ov p$ on protons and 
%deuterons between 50 and 200 GeV/c)
 Phys. Rev. Lett. 33 (1974) 928;
%(Total cross sections of $\pi^{+}, \pi^{-}$, $K^{+}, K^{-}$ on p and d between
% 50 and 200 GeV/c)
 Phys. Rev. Lett. 33 (1974) 932;
%(Total cross sections of $\pi^{+}, \pi^{-}$,
% $K^{+}, K^{-}, p$ and $\ov p$ from 23 to 280 GeV/c) 
Phys. Lett. 61B (1976) 303;
%(Total cross sections of $\pi^{+}, \pi^{-}, K^{+}, K^{-}, p$ and $\ov p$ and 
%d between 200 and 370 GeV/c)
 Phys. Lett. 80B (1979) 423.} \vspace{-0.3cm}
\item{A. S. Carroll et al.,
% (Absorption  cross sections of $\pi^{+}, \pi^{-}, 
%K^{+}, K^{-}, p$ and $\ov p$ on nuclei between 60 and 280 GeV/c)
 Phys. Lett. 80B (1979) 319.} \vspace{-0.3cm}
\item{G. Giacomelli,
% (Total cross section measurements)
 Progress in Nuclear 
Physics 12 (1970) 77.} \vspace{-0.3cm}
\item{ W. Bozzoli et al.,  Nucl. Phys. B140 (1978) 271;
Nucl. Phys. B144 (1978) 317; Nucl. Phys. B159 (1979) 363;
A. Bussiere et al.,
Nucl. Phys. B174 (1980) 1.} \vspace{-0.3cm}
\item{U. Amaldi et al.,
% (The energy dependance of the pp total cross section for
%c.m. energies between 23 and 53 GeV)
 Phys. Lett. 44B (1973) 112;
S. R. Amendolia et al.,
% (Measurments of the total pp cross sections at the ISR),
Phys. Lett. 44B (1973) 119;
G. Carboni et al.,
% (Precise measurements of $\ov p p$ and $pp$ total cross
% section at the CERN-ISR)
 Nucl. Phys. B254 (1985)69;
G. Giacomelli and M. Jacob, 
%(Physics at the CERN-ISR)
 Phys. Rep. 55 (1979) 1.} \vspace{-0.3cm}
\item{A. Bertin et al., Phys. Lett. 38B (1972) 260;
Phys. Lett. 42B (1972) 493; M. Antinucci et al.,
Nuovo Cimento Lett. 6 (1973) 121;
P. Capiluppi et al., Nucl. Phys. B70 (1974) 1;
Nucl. Phys. B79 (1974) 189;
E. Albini et al.,
Nuovo Cimento 32 (1976) 101.} \vspace{-0.3cm}
\item{A. Breakstone et al.,
Phys. Lett. 132B (1983) 458;
Nucl. Phys. B248 (1984) 253;
Phys. Rev. Lett. 54 (1985) 2180.} \vspace{-0.3cm}
\item{N. A. Amos et al.,
% (Measurement of the $\ov p p$ total cross sections at
% $\sqrt s =$ 1.8 TeV)
 Phys. Rev. Lett. 63 (1989) 2784;
% (Antiproton-proton 
%elasctic scattering at $\sqrt s =$ 1.8 TeV from -t=0.034 to 0.65 (GeV/c)**2)
Phys. Lett. B247 (1990 127; 
%(A luminosity-indipendent measurement of the $\ov p p
%$ total cross section at $\sqrt s =$ 1.8 TeV )
Phys. Lett. B243 (1990)158.} \vspace{-0.3cm}
\item{W. F. Baker et al., (Production of $\pi^{\pm}, K^{\pm},  p$ and $\ov p$ by
400 GeV/c protons) Fermilab-78/79-EXP (1978).} \vspace{-0.3cm}
\item{ G. Giacomelli,
Int. J. Mod. Phys. A5 (1990) 223; Hadron-hadron and hadron-nuclei interactions at intermediate and
high energies 1994 Marshak Memorial, DFUB 9/94 (1994);
 Total cross sections, Kycia Memorial Symposium, 19/5/2000, BNL,
hep-ex/0006038; Progress in Nuclear Physic  s 12 (1970) 77.} \vspace{-0.3cm}
\item{S. Cecchini et al.,
Astrop. Phys. 1 (1993) 369; G. Giacomelli et  al., Nucl. Instr. Meth. A411 (1998)
 41; H. Dekhissi et al., Nucl. Phys. A662 (2000) 207.} \vspace{-0.3cm}
\item{A. Capella et al., Phys. Rep. 236 (1994); J. J. Glauber et al., Nucl.
 Phys. B21 (1970) 135; J. Ranft, Phys. Rev. D51 (1995) 64.}

\end{enumerate}
\end{document}